\documentclass[english,a4paper]{article}
\usepackage{dcolumn}
\usepackage{bm}
\usepackage{graphicx}
\usepackage{amsmath}
\usepackage{amsfonts}
\usepackage{amssymb}
\usepackage{amsthm}
\usepackage{amstext}
\usepackage{amsbsy}
\usepackage{amsopn}
\usepackage{amscd}
\usepackage{amsxtra}

\def\phi{\varphi}
\def\epsilon{\varepsilon}

\everymath{\displaystyle}

\begin{document}
\title{Characterization of a Driven Two-Level Quantum System by~Supervised Learning}
\author{R. Couturier\footnote{Universit\'e de Franche-Comt\'e, CNRS, Institut FEMTO-ST, F-90000 Belfort, France}, E. Dionis\footnote{Laboratoire Interdisciplinaire Carnot de
Bourgogne (ICB), UMR 6303 CNRS-Universit\'e Bourgogne, 9 Av. A.
Savary, BP 47 870, F-21078 Dijon Cedex, France}, C. Guyeux\footnote{Universit\'e de Franche-Comt\'e, CNRS, Institut FEMTO-ST, F-90000 Belfort, France}, S. Gu\'erin\footnote{Laboratoire Interdisciplinaire Carnot de
Bourgogne (ICB), UMR 6303 CNRS-Universit\'e Bourgogne, 9 Av. A.
Savary, BP 47 870, F-21078 Dijon Cedex, France},D. Sugny\footnote{Laboratoire Interdisciplinaire Carnot de
Bourgogne (ICB), UMR 6303 CNRS-Universit\'e Bourgogne, 9 Av. A.
Savary, BP 47 870, F-21078 Dijon Cedex, France, dominique.sugny@u-bourgogne.fr}}

\maketitle

\begin{abstract}
We investigate the extent to which a two-level quantum system subjected to an external time-dependent drive can be characterized by supervised learning. We apply this approach to the case of bang-bang control and the estimation of the offset and the final distance to a given target state. For any control protocol, the goal is to find the mapping between the offset and the distance. This mapping is interpolated using a neural network. The estimate is global in the sense that no a priori knowledge is required on the relation to be determined. Different neural network algorithms are tested on a series of data sets. We show that the mapping can be reproduced with very high precision in the direct case when the offset is known, while obstacles appear in the indirect case starting from the distance to the target. We point out the limits of the estimation procedure with respect to the properties of the mapping to be interpolated. We discuss the physical relevance of the different results.
\end{abstract}

\section{Introduction}
Machine learning is a field of computer science which has attracted recently much attention in many areas in physics~\cite{carleoRMP2019,mehtaPhysRep2019,judsonPRL1992}. The algorithms are aimed to emulate human intelligence by learning the best way to proceed from a large data set~\cite{mnihNature2015}. The power of this tool gives hope that long outstanding problems can be solved in quantum physics~\cite{HushScience2017,dunjkoQV2020}. Pioneering studies have applied such techniques with success in various domains~\cite{krenn2023} extending from many-body physics~\cite{CarleoScience2017,carrasquillaNatPhys2017} to quantum computing~\cite{carleoRMP2019,dunjkoPRL2016,dingPRL2020}. Different problems can be tackled from machine learning techniques. They are usually classified into three categories, namely supervised learning (SL), unsupervised learning, and reinforcement learning (RL)~\cite{carleoRMP2019}. As illustrative examples, different results have been obtained very recently. Latent neural ordinary differential equations have been used to learn, reconstruct, and extrapolate quantum trajectories with high accuracy~\cite{Choi2022}. A stochastic learning approach has been applied for controlling open quantum systems~\cite{YY2022}. A machine learning algorithm has been proposed for continuous quantum error correction~\cite{Convy2022}. On another side, the control of quantum systems by means of intense electromagnetic pulses has been a topic of increasing
interest in the past decades~\cite{cat,dalessandro-book,past-present-future,STAreview,odelin2019} with a variety of applications extending from atomic and molecular physics~\cite{past-present-future,RMPsugny} to magnetic resonance~\cite{cat} and quantum technologies~\cite{roadmapQT,kochroadmap}. The recent progress of numerical optimization techniques and experimental devices has made possible the design and the implementation of controls able to manipulate with precision quantum systems of growing complexity. In this setting, reinforcement learning can be used to solve such issues~\cite{giannelli2022,martin2021}. RL differs from other types of machine learning in that the system is not trained with an example data set. Instead, the system learns through a trial and error method. This approach has been explored recently in benchmark quantum control problems with success (see Refs.~\cite{bukovPRX2018,dayPRL2019,zhang2018,An_2019,porotti2019,niu2019,wangPRL2020,ding2021,borah2021,yao2021,chen2022,sgroi2021,Brown2021,Cao2022,erdman2022} to mention a few). Note that robust quantum controls with respect to small system uncertainties can also be designed when formulated as a SL task~\cite{wuPRA2019,yang2018,kestner22}. Convolutional neural networks and SL technique have been used recently to design control protocols in a random environment~\cite{Huand2022}.

Despite recent and impressive success, such open-loop control methods have intrinsic limitations when they are implemented in a realistic experimental setting. Among others, they require the accurate knowledge of the system dynamics  and a quick estimate of the efficiency of the considered control process~\cite{roadmapQT,kochroadmap,wittler2021}. Estimation of system parameters in quantum control has recently been widely investigated using different inversion techniques~\cite{lebris,geremia03,geremia2002b,Ndong2014,schirmer2015,zhang2014,wittler2021,sone2017,burgarth2017,xue2021,buchwald2021} based, e.g., on quantum Fisher information~\cite{haidong2015,haidong2017,lin2021,lin2022,yang22} or fingerprinting approaches~\cite{mri2013,ansel2017}. {Parameter estimation can also be enhanced by machine learning methods as shown, e.g., in~\cite{lumino2018,xiao2022}.} In spite of their efficiency, these methods can be viewed generally as local since they require a good estimate of the parameter to be determined. This constraint can be partly avoided by using machine learning techniques such as SL and neural network (NN) algorithms in which the estimation may be global without any prior knowledge or at least with a minimum information on the value to find~\cite{hopfield1988artificial,pedregosa2011scikit}. This latter approach has been applied successfully in recent examples to, e.g., extract the noise spectrum from system dynamics~\cite{papi22,wise21} or to characterize the non Markovianity of open quantum systems~\cite{fanchini21}. Similar techniques have also been used to identify quantum system Hamiltonian~\cite{geremia2002}.

In this framework, this paper aims to identify the obstacles to the application of SL in the estimation of parameters of quantum dynamics. In this study, we focus specifically on the implementation of SL to the characterization of driven two-level quantum systems. In order to highlight the advantages and the fundamental limitations of this technique, we consider a minimal, but non-trivial, model involving a two-level quantum system subjected to a non-resonant real control~\cite{boscain-mason,sugny10,boscainPRXQ2021}, in which the offset of the system is denoted $\Delta$. This reference system is well-known in quantum control~\cite{roadmapQT,kochroadmap}. For instance, a time-optimal solution to reach a given target state can be derived if the maximum intensity of the external field is bounded~\cite{boscain-mason,sugny10,boscainPRXQ2021}. The optimal solution to steer the system from the ground to the excited state is a bang-bang pulse, that is a pulse of maximum intensity with a switch from the positive to the negative amplitude at a specific time of the control process. Inspired by this procedure, we study in this paper a similar control process and we assume that the external control can switch a finite number of times (fixed to five at random times in the numerical simulations) between its maximum and minimum values during a given control duration, corresponding to the previous minimum time. As in a standard control problem, the initial and target states are the ground and excited states of the system and we define a distance $d$ between the final dynamical state and the target.

In a SL process, the goal is to find a mapping able through a suited neural network to associate a set of inputs to outputs. SL is roughly divided into two stages. The first step is a learning procedure where the parameters of the NN are optimized based on an input--output data set. In a second time, another set of data is used to test the precision of the NN to reproduce the targeted mapping. If the tests are conclusive then the NN becomes a very powerful tool allowing in a very short time to determine from any input the corresponding output. This very attractive procedure for dynamical systems nevertheless presents difficulties and limits. To this aim, we apply this general framework to two different characterization processes. In the first case, knowing the control and the offset, the goal is to find the distance to the target state, while the role of $\Delta$ and $d$ is reversed in the second analysis. As discussed below, the first and second SL issues can be used, respectively, to characterize either the final state of the system or one of the parameters of the Hamiltonian. They will be called below \emph{{direct}} and \emph{{inverse}} estimation problems. On the basis of large data sets, we investigate on this two fundamental example the connection between the complexity of the NN and the accuracy of the SL. We show that intrinsic limits to the precision of this process exist and we quantify them for this model system. We point out qualitative characteristics that the mapping must verify to be well reproduced by a NN. The results are established for a specific control of a two-level quantum system but the conclusions obtained in this study can be generalized to the application of SL to other quantum dynamical processes.

The remainder of this paper is organized as follows. Section~\ref{sec2} introduces the physical model and describes the time-optimal solution for steering the quantum system from the ground state to the excited one. The principles of supervised learning are outlined in Section~\ref{sec3}, with special attention paid to its application in quantum control. The numerical results are presented and discussed in Section~\ref{sec:numres}. We conclude in Section~\ref{sec5} with an outlook. Additional material is provided in Appendix~\ref{appa}.

\section{The Model System}\label{sec2}
This section aims at describing the model system under consideration and the known results about its optimal control.

We consider the control of a two-level quantum system by means of an external electromagnetic field. The state of the system at time $t$ is $\psi(t)\in\mathbb{C}^2$ of coordinates $(c_1,c_2)$ and the norm of $\psi$ is equal to 1. The dynamics are governed by the time-dependent Schr\"odinger equation $i\dot{\psi}=H\psi$
in units where $\hbar=1$, $H$ being the Hermitian Hamiltonian matrix. In a given rotating frame and in the rotating wave approximation, this latter can be expressed as
$$
H=\frac{1}{2}\begin{pmatrix}\Delta & u\\
u & -\Delta\end{pmatrix}
$$
with $\Delta\in\mathbb{R}$ is the offset term with respect to the frequency of the field and $u(t)\in\mathbb{R}$ the control law, that corresponds to the amplitude of the excitation. The goal of the optimal control problem is to steer in minimum time the system from the ground to the excited state, i.e., to go from the initial state $\psi_0=(1,0)$ to the target $\psi_f=(0,1)$ (up to a phase factor).

This control problem can be reformulated in real coordinates by introducing the following change of coordinates
$$
\begin{cases}
x=c_1c_2^*+c_1^*c_2 \\
y=-i(c_1c_2^*-c_1^*c_2) \\
z=|c_1|^2-|c_2|^2
\end{cases}
$$
with the constraint $x^2+y^2+z^2=1$, which corresponds to the Bloch sphere. The dynamical system can then be expressed as
\begin{equation}\label{eqbloch}
\begin{cases}
\dot{x}=-\Delta y\\
\dot{y}=\Delta x-uz \\
\dot{z}=uy
\end{cases}
\end{equation}
The goal is now to bring the system from the north pole ($z=1$) to the south pole ($z=-1$) of the sphere. We add a pulse limitation $|u(t)|\leq u_0$ to the control, for some $u_0>0$. Note that a time rescaling leads to the multiplication of $\Delta$ and $u_0$ by a positive scalar, and to the normalization $u_0=1$.

The dynamical system on the Bloch sphere can be exactly integrated for a bang control for which $u=\pm 1=\varepsilon$. We assume that the initial point is $(x_0,y_0,z_0)$ at $t=0$. Using
$$
\ddot{y}=\Delta \dot{x}-\varepsilon\dot{z}=-(1+\Delta^2)y,
$$
which leads to $\ddot{y}+\Omega^2 y=0$, with $\Omega=\sqrt{1+\Delta^2}$, we deduce that
$
y(t)=A\cos(\Omega t)+B\sin(\Omega t).
$
Since $y(0)=y_0$, we have $A=y_0$. With $\dot{y}(0)=\Delta x_0-\varepsilon z_0$, we deduce that
$
B=\frac{\Delta x_0-\varepsilon z_0}{\Omega} .
$
For the $x$- and $z$- coordinates, we have\vspace{-3pt}
$$
\begin{cases}
x(t)=x_0-\frac{\Delta B}{\Omega}-\frac{\Delta}{\Omega}(A\sin(\Omega t)-B\cos(\Omega t)) \\
z(t)=z_0+\frac{\varepsilon B}{\Omega}+\frac{\varepsilon}{\Omega}(A\sin(\Omega t)-B\cos(\Omega t))
\end{cases}
$$

For a specific value of $\Delta$, this problem can be solved explicitly by optimal control and the Pontryagin Maximum Principle. We refer the interested reader to~\cite{boscain-mason,sugny10,boscainPRXQ2021} for details on the derivation of the optimal solution. The minimum time $t^*$ to solve the control problem can also be found. For $\Delta\leq 1$, it can be shown that the optimal solution is the concatenation of two bang arcs of amplitude $\pm 1$.

\begin{figure}[tb]
\centering
\includegraphics[scale=0.5]{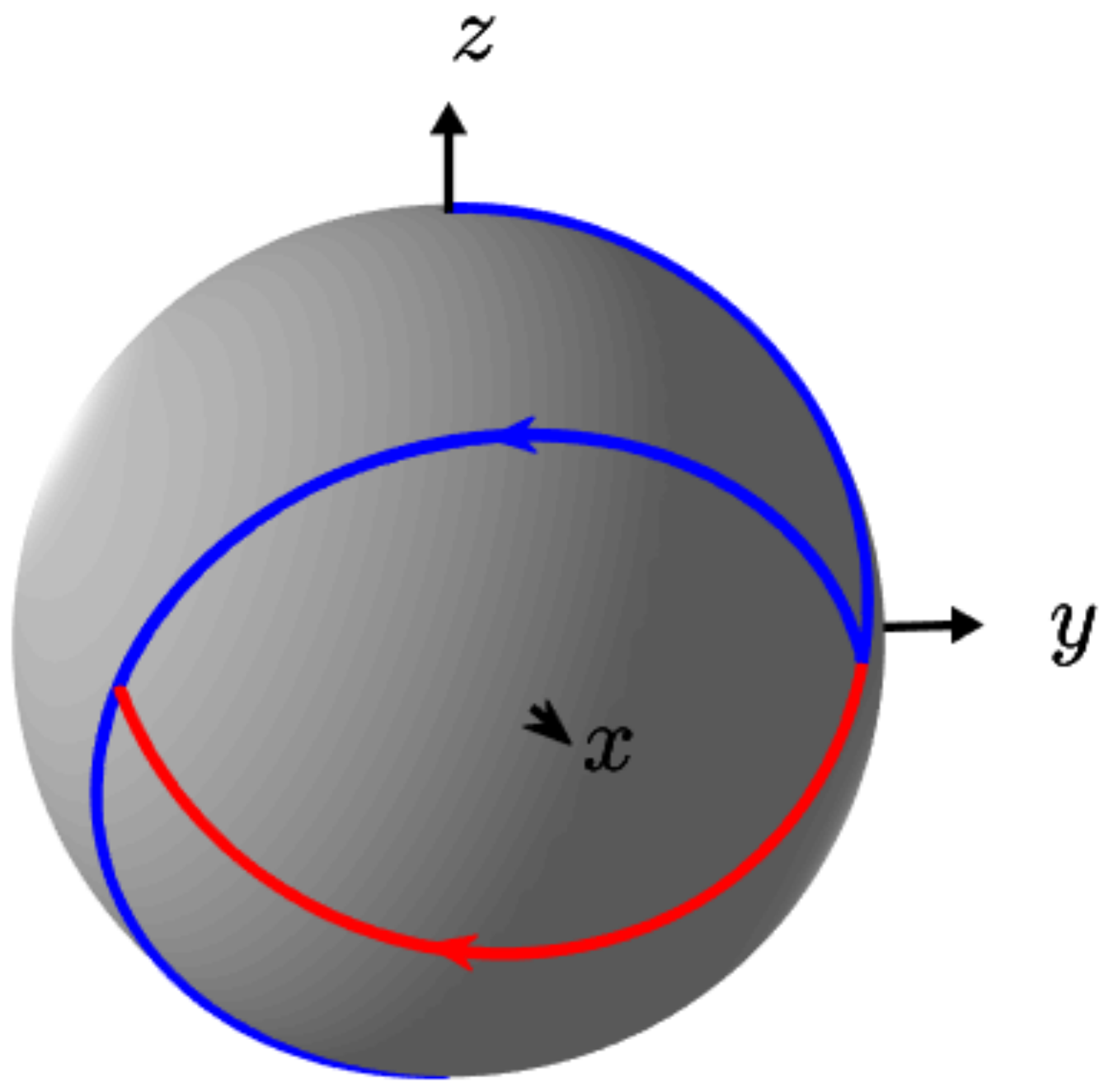}
   \caption{Optimal trajectories (in blue) going from the north pole to the south pole
of the Bloch sphere. The parameter $\Delta$ is set
to -0.5. For one of the two optimal trajectories a control $u=-1$ is first applied during a time $t_1$, followed by a control $u=+1$ during a time $t_2$, while
for the other trajectory $u=-1$ lasts for a time $t_2$ and $u=+1$ for a time $t_1$ (the specific part of the second solution is plotted in red).\label{fig1}}
  \end{figure}

As displayed in Figure~\ref{fig1}, the control sequence is characterized by two times $t_1$ and $t_2$ defined as:
$$
\begin{cases}
t_1=\frac{1}{\Omega}(\pi-\arccos(\Delta^2)) \\
t_2=\frac{1}{\Omega}(\pi+\arccos(\Delta^2)) \\
\end{cases}
$$
with $\Omega=\sqrt{1+\Delta^2}$ and $t^*=\frac{2\pi}{\Omega}$. Note that there are two symmetric time-optimal solutions. Inspired by this control protocol, we consider below to characterize the quantum system a similar bang-bang control sequence with five switches and a time fixed to $t^*$.

\section{Methodology}\label{sec3}
\subsection{Principles of Machine Learning Techniques}

Machine learning is a form of artificial intelligence that allows a system to learn from data, not through explicit programming. A machine learning model is the result generated when the machine learning algorithm is trained with data. After training, when a model receives data as input, it produces a prediction as output. As mentioned earlier, there are several types of machine learning processes.

Unsupervised learning is used when the problem requires a massive amount of unlabeled data. To understand the meaning of this data, it is necessary to use algorithms that classify the data according to the patterns or clusters they detect. Supervised learning, on the other hand, typically begins with a well-defined data set and some understanding of how that data is classified. The goal of SL is to detect patterns in the data and apply them to an analytical process. These data have features associated with labels that define their meaning.
Deep learning, finally, is a specific method of SL that integrates neural networks in successive layers to learn data in an iterative way~\cite{lecun2015deep}. 
The purpose of the present paper is to show how to apply and to test the efficiency of SL and of NN algorithms for the characterization of the driven two-level quantum system described in Section~\ref{sec2}.
\subsection{Construction of an Artificial Neural Network}
The goal of this paragraph is to illustrate the construction of a NN in its basic form of a multilayer perceptron (MLP). To this aim, we consider the following global characterization problem for the two-level quantum system of Section~\ref{sec2}. Given the offset $\Delta$ and the control law $u(t)$, the goal is to find the distance to the south pole of the Bloch sphere starting from the north pole. An exact transfer from the north to the south pole is given in Figure~\ref{fig1}. To limit the complexity of the learning process, we assume that $\Delta\in [0,1]$ and that $u$ is a piecewise constant function of 100 equal steps, $u_k$ of values 1 or $-$1 with exactly 5 switches. The control time is fixed to $T=t^*$ (note that this duration depends on the detuning $\Delta$). Based on this simple but fundamental control problem, our aim is to propose a robust learning process and evaluate the relative effectiveness of different supervised learning techniques. We describe in details in this paragraph the different steps of the application of such algorithms. Other examples are studied in Section~\ref{sec:numres}.

We consider the following data set. We draw randomly 10 offsets in $[0,1]$ and 10 million  controls for each offset. This corresponds to a vector denoted $X$ with 101 entries (one offset and 100 values $u_k$). For each vector, we numerically compute the final state $(x(T),y(T),z(T))$ by a direct integration of Equation~\eqref{eqbloch}. We obtain its Euclidean distance to the target defined as $Y=\sqrt{x(T)^2+y(T)^2+(z(T)+1)^2}$. The data set consisting of 10 million elements with input $X$ and output $Y$ is separated into 80\% for the training and 20\% for the testing processes. The second step of this approach consists of building a MLP network able to estimate the output from the given of the input.

\begin{figure}[tb]
    \centering
    \includegraphics[scale=0.3]{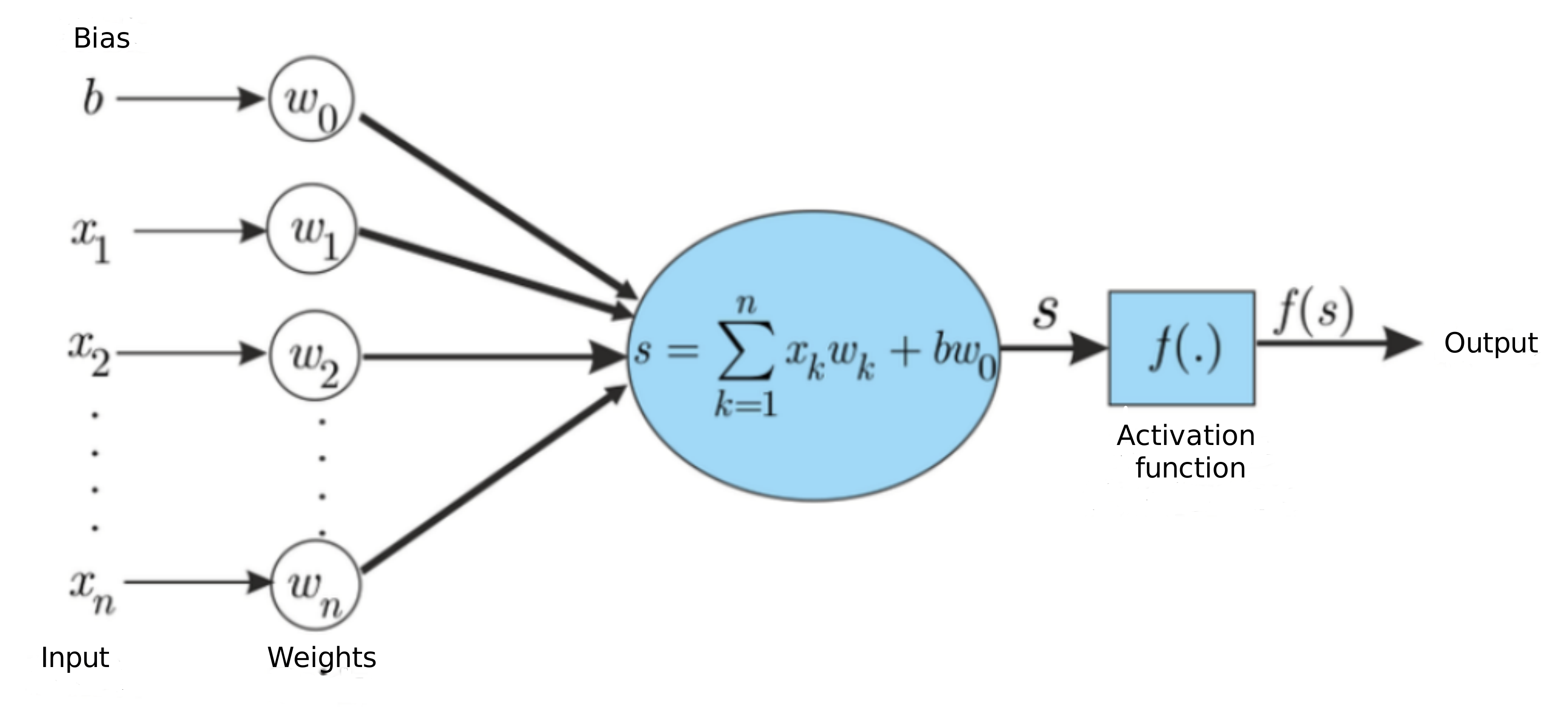}
    \caption{Schematic representation of an artificial neuron used in the numerical simulations.}
    \label{fig:neurone}
\end{figure}

We first recall the functioning of an artificial NN in its basic form of a MLP. Its basic component is the artificial neuron, as detailed in Figure~\ref{fig:neurone}. In this computation unit, a linear combination of its weighted input is evaluated, and a weighted bias (a real number) is added. In the different numerical simulations, the bias number $b$ is set to 1 and the parameter $n$ to 101. An activation function $f$, which introduces non-linearity to the estimation problem, is applied to this value, where $f$ is chosen among a small list of usual functions, extending from a threshold function, the sigmoid to the tanh functions or even a ReLu function (rectified linear activation function). Finally, the obtained result is published as output of the artificial neuron. Notice that for each neuron, $n+1$ weights have to be chosen in order to define the mapping associating the input to the output.

\begin{figure}[tb]
    \centering
    \includegraphics[scale=0.3]{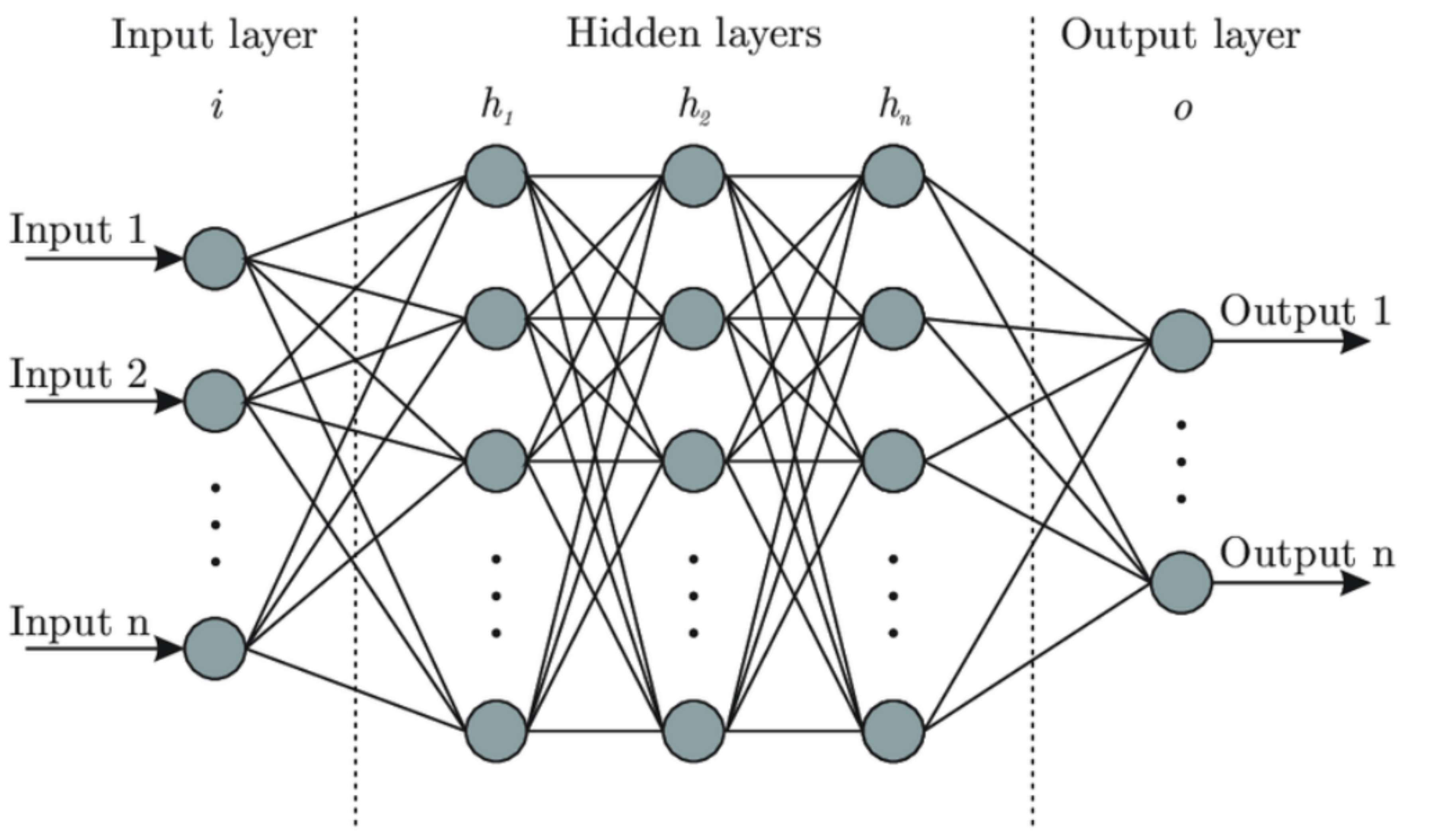}
    \caption{Description of the architecture of a multilayer perceptron. Each gray disk represents an artificial neuron as illustrated in Figure~\ref{fig:neurone}. The neurons are organized in columns, each corresponding to a layer, and called input, hidden, or output layers according to their position in the structure of the MLP. There are $n$ neurons by layer. Their output is used as input for the neurons of the next layer as represented by the solid lines.}
    \label{fig:MLP}
\end{figure}

These basic computation units are grouped by layers in the MLP architecture, as depicted in Figure~\ref{fig:MLP}. Each neuron of each internal layer is connected to all the neurons of the previous layer. As can be seen in Figure~\ref{fig:MLP}, we distinguish between the input layer which receives the data and transfers it unchanged to the first hidden layer. Each hidden layer contains $n$ neurons. The objective is then to find the best weights and biases that most closely coincides with a given basis of knowledge that correlates in a univocal way a set of input to a set of output. More precisely, it must find the best weights that fit on a ``training set'' (80\% of the basis of knowledge), such that the ``predictions'' in the remained test set are the closest to reality. The weights are found through a specific optimization process called gradient backpropagation. For our problem, the MLP input is constituted of 101 inputs,  corresponding to the 100 values of the control and of the offset $\Delta$. At the final stage, there is only one output, namely the distance $Y$. Additional details about the optimization of the different free parameters of the MLP are given in Section~\ref{sec:numres}.

\vspace{-12pt}

\section{Numerical Results}\label{sec:numres}

Intensive numerical simulations have been performed to estimate the efficiency of different supervising learning approaches in this fundamental control problem. We have determined four data sets denoted below (1), (2), (3), and (4). The data set (1) is the one described in Section~\ref{sec3} for a total time $T=t^*$. The data set (3) is the same as (1), except that 100 offsets uniformly distributed in $[0,1]$ have been used. In the data sets (2) and (4), we consider, respectively, the same control fields as in (1) and (3), but we exchange the role of the distance and the offset. The goal of the process is now to estimate the value of $\Delta$, which corresponds to the output $Y$ of the neural network. Thus, this is a supervised learning problem, where the label of the synthetic basis of knowledge is either the distance to the target state or the $\Delta$ parameter, depending on the dataset. Note that the training and testing samples are identically and independently distributed, as $\Delta$ (or the distance in the reverse problem) and the steps have been chosen as random.

Several NN algorithms were first tested on data set (1). The best architecture was then applied to the other data sets, by playing on the hyperparameters of the model. Notice that we do not claim to have found the best structure for the NN. An exhaustive search that requires a significant amount of time is beyond the scope of this study. However, the tested architectures lead to key conclusions about the efficiency and choice of methods.

\begin{figure}[tb]
\centering
\includegraphics[scale=0.5]{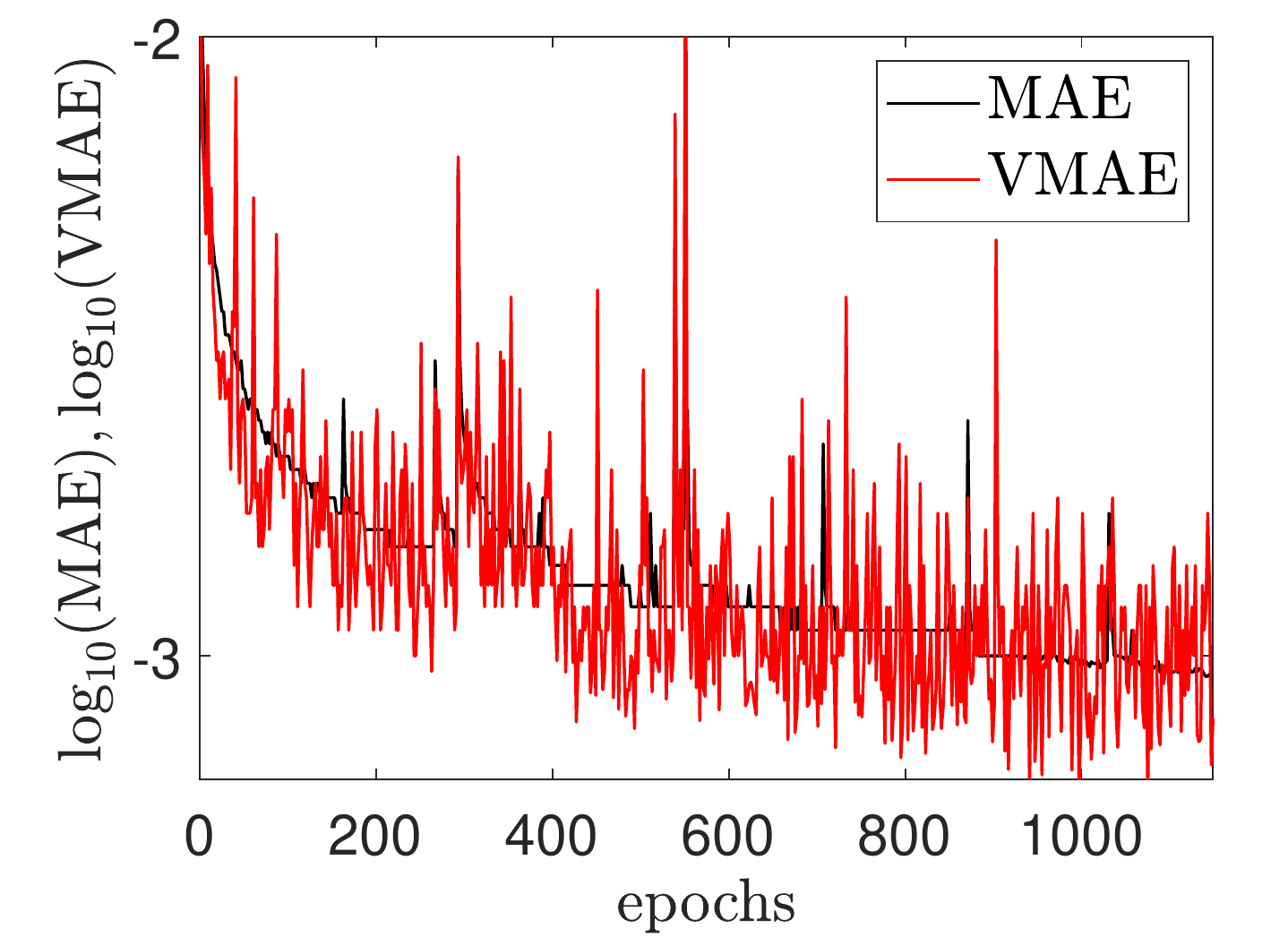}
   \caption{Evaluation 
 of the training mean average error (MAE) and testing (VMAE) during the optimization of the MLP NN with data set 2 (see the text for details).\label{fig:mae}}
  \end{figure}

A MLP process is first investigated. This architecture is clearly the simplest one for regression. Weights have been initialized by the He uniform kernel initializer, while all activation functions have been set to the ReLU one. Indeed, ReLU advantages are sparsity and a reduced likelihood of vanishing gradient. Additionally, deep models have difficulties in converging when weights are normalized with fixed standard deviation, leading to large or small activation values which result in exploding or vanishing gradient when backpropagation. This problem is overcome thanks to the He initializer, which takes into account the non-linearity of ReLU activation functions. The last layer corresponding to the output one is made of one neuron, as we want to predict a single value (the distance) and the ReLU activation function, as all neurons must have a non-linear character. The Adam optimizer and the mean squared error as loss function have been used. The number of epochs has been limited to 1000 with an early stopping criteria on a validation data set (20\% of the training data set), and the batch size has been set to 128, to reduce the memory consumption and improve the convergence speed (too small values of batch size take long time to converge and too large ones may converge less well). Finally, for sake of completeness, other optimizers, kernel initializers, and activation functions have been tested together with other models (namely, GRU and transformer-based ones), but without improving the test scores. The optimal number of layers has also been studied. A good compromise is to use 14 layers, for the first half of the layers, the number of neurons is doubled at each layer where for the second half the number of neurons is divided by two at each layer.
The second family of architectures is based on Convolutional Neural Networks (CNN), with one (or more) couple(s) of Convolution 1D layer(s). After some layers, neurons are flatten and a small MLP is used. Based on the previous computations, a ReLU function has been chosen as activation in the convolutional and dense layers, with the He uniform initializer and the Adam optimizer. A good compromise is to use seven layers of convolution size three. For each new layer, the number of filters is doubled. Then a MLP is used with seven layers. Finally, a stacked Long Short-Term Memory (LSTM) approach has been tested on the first data set. This recurrent neural network approach is promising, because of the particular structure of the control field, and the recurrent effect of each switch on the final distance. Four stacked LSTMs are used. The neural architecture was finalized with a small dense hidden layer (ReLU activation). Finally, at each time, the Keras default loss function has been chosen, namely the mean squared error. Figure~\ref{fig:mae} represents the training and testing errors with the Mean Average Error (denoted, respectively, MAE and VMAE) for the training and testing data sets. In this figure, it can be seen that the MAE decreases rapidly for the first one hundred epochs and then slowly afterwards. As could be expected, the testing MAE (VMAE) exhibits a more oscillating behavior since the hyperparameters are not optimized on this part of the data set.

Another numerical characteristic is explored in Table~\ref{tabnew}. We study for a specific dataset how the precision of the NN evolves according to the number of elements. We observe that a relatively large data set is needed to achieve good efficiency.

The different numerical results for the data sets (1) and (2) are given in Table~\ref{tab:dataset12}. We observe for the example (1) that the results are quite good and lower than $10^{-3}$ for all the architectures. Table~\ref{tab:dataset12} also shows that very different NN architectures lead more or less to similar results. Due to the huge number of data, the time to train the different architectures is not negligible and of the order of one day in each case (with a Nvidia V100 GPU). The results achieved for the data set (2) are clearly different because the obtained MAE is of the order of $10^{-2}$, i.e., two times larger than in the first case, while the estimation conditions seem at first sight to be very similar.  None of the tested NN architecture were able to solve this problem with a sufficient efficiency. Here again, we point out that all the tested NN in spite of their different complexity leads to equivalent results.

\begin{table}[tb]
\centering
\caption{Evolution of the testing mean average error (VMAE) as a function of the number of elements in the data set. The first column indicates the fraction of the maximum number of elements used in the learning process, which is fixed to 10 million. Numerical values have been computed for the data set 1 with the LSTM NN (see the text for details).}
    \begin{tabular}{cc}
    \hline
    \textbf{Fraction} & \textbf{VMAE} \\
    \hline
    \hline
     1 & $2.78\times10^{-4}$ \\
    \hline
    1/2 &  $3.93\times10^{-4}$ \\
    \hline
    1/4 & $6.13\times10^{-4}$ \\
    \hline
    1/8 & $1.1\times10^{-3}$ \\
   \hline
    \end{tabular}

    \label{tabnew}
\end{table}


\begin{table}[tb]
    \caption{Results of the different architecture for the data sets 1 and 2 corresponding to the second and third columns of the table. The MLP, CNN, and LSTM algorithms are used to estimate the output. VMAE is the mean average error between the exact output values and the estimations for the testing data set. }
    \centering
    \begin{tabular}{ccc}
    \hline
    \textbf{Algorithm} & \textbf{VMAE (1}) & \textbf{VMAE (2)} \\
    \hline
    \hline
     MLP & $5.35\times10^{-4}$ & $5.16\times10^{-2}$ \\
    \hline
    CNN &  $8.50\times10^{-4}$ & $5.36\times10^{-2}$\\
    \hline
    LSTM & $2.78\times10^{-4}$ & $5.00\times10^{-2}$ \\
    \hline
    \end{tabular}

    \label{tab:dataset12}
\end{table}

Numerical simulations on data sets (3) and (4) are summarized in Table~\ref{tab:dataset34}, while this estimation process seems a bit more difficult than the first case, surprisingly slightly better results are observed for the different NN. From an algorithmic point of view, this means that slightly more accurate estimations can be made. This could be explained by the fact that CNNs and LSTMs take into account the temporal side of the application of switches $u_k$, through convolution for the former, and the recurrent character for the latter. Similarly to the data set (2), NN are not able to find the right values of the offset $\Delta$ for the set (4).

\begin{table}[tb]
 \caption{Same as~\ref{tab:dataset12} but for the data sets 3 and 4.}
    \centering
    \begin{tabular}{ccc}
    \hline
    \textbf{Algorithm} & \textbf{VMAE (3)} & \textbf{VMAE (4)} \\
    \hline
    \hline
     MLP & $3.03\times10^{-4}$ & $7.95\times10^{-2}$ \\
    \hline
    CNN &  $2.78\times10^{-4}$ & $6.86\times10^{-2}$\\
    \hline
    LSTM & $1.76\times10^{-4}$ & $6.26\times10^{-2}$ \\
    \hline
    \end{tabular}

    \label{tab:dataset34}
\end{table}

\begin{figure}[tb]
\centering
\includegraphics[scale=0.5]{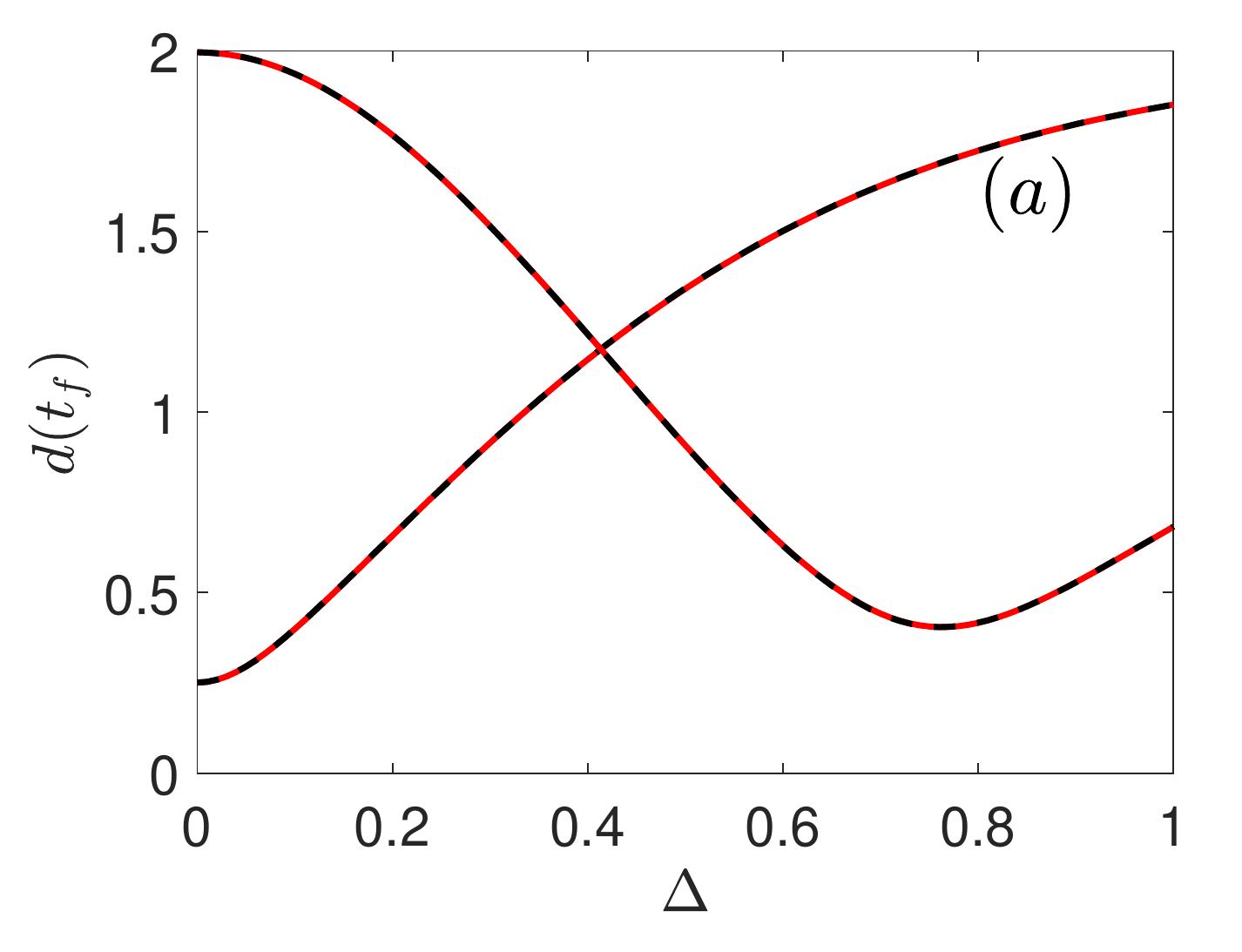}\\
\includegraphics[scale=0.5]{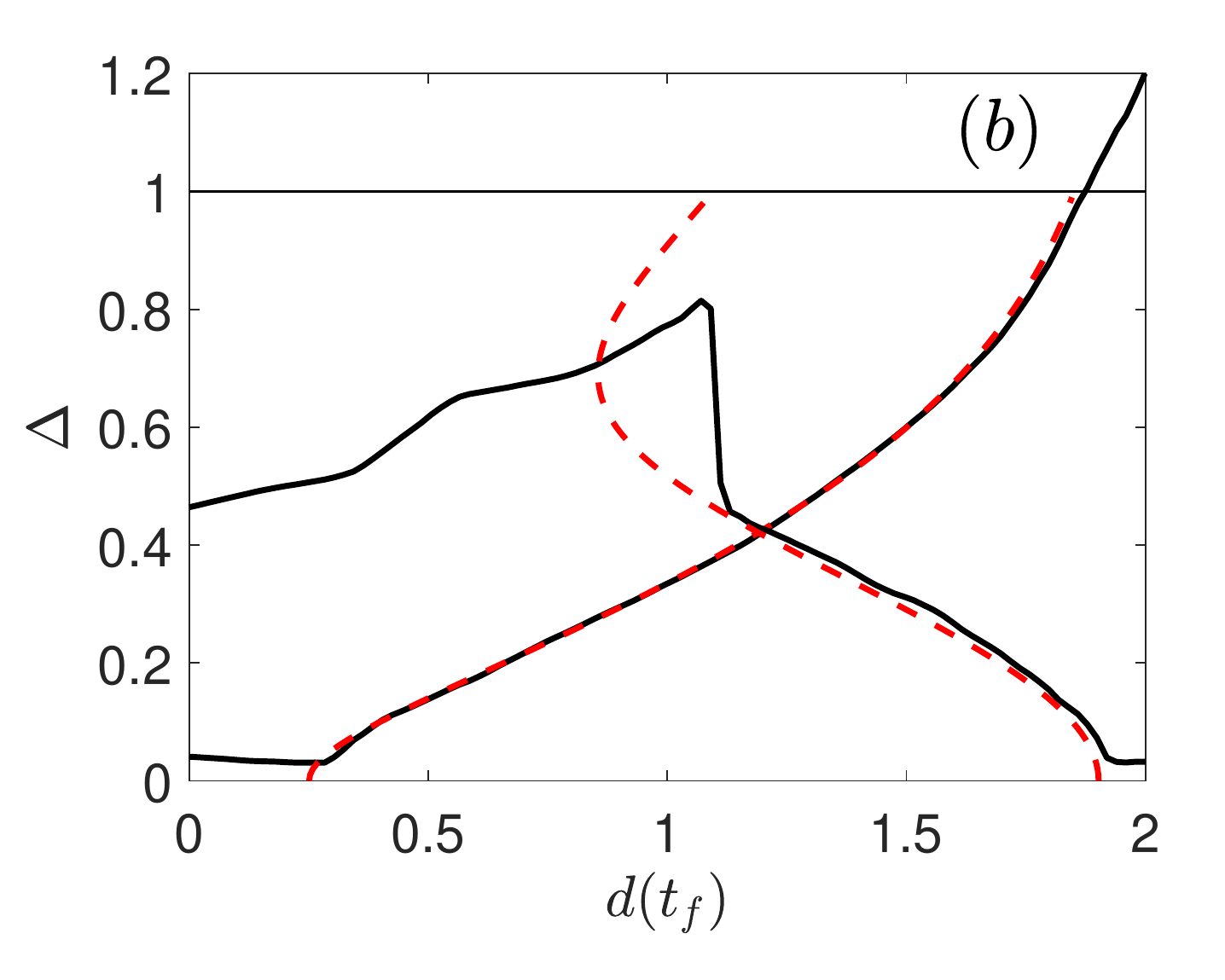}\\
\includegraphics[scale=0.5]{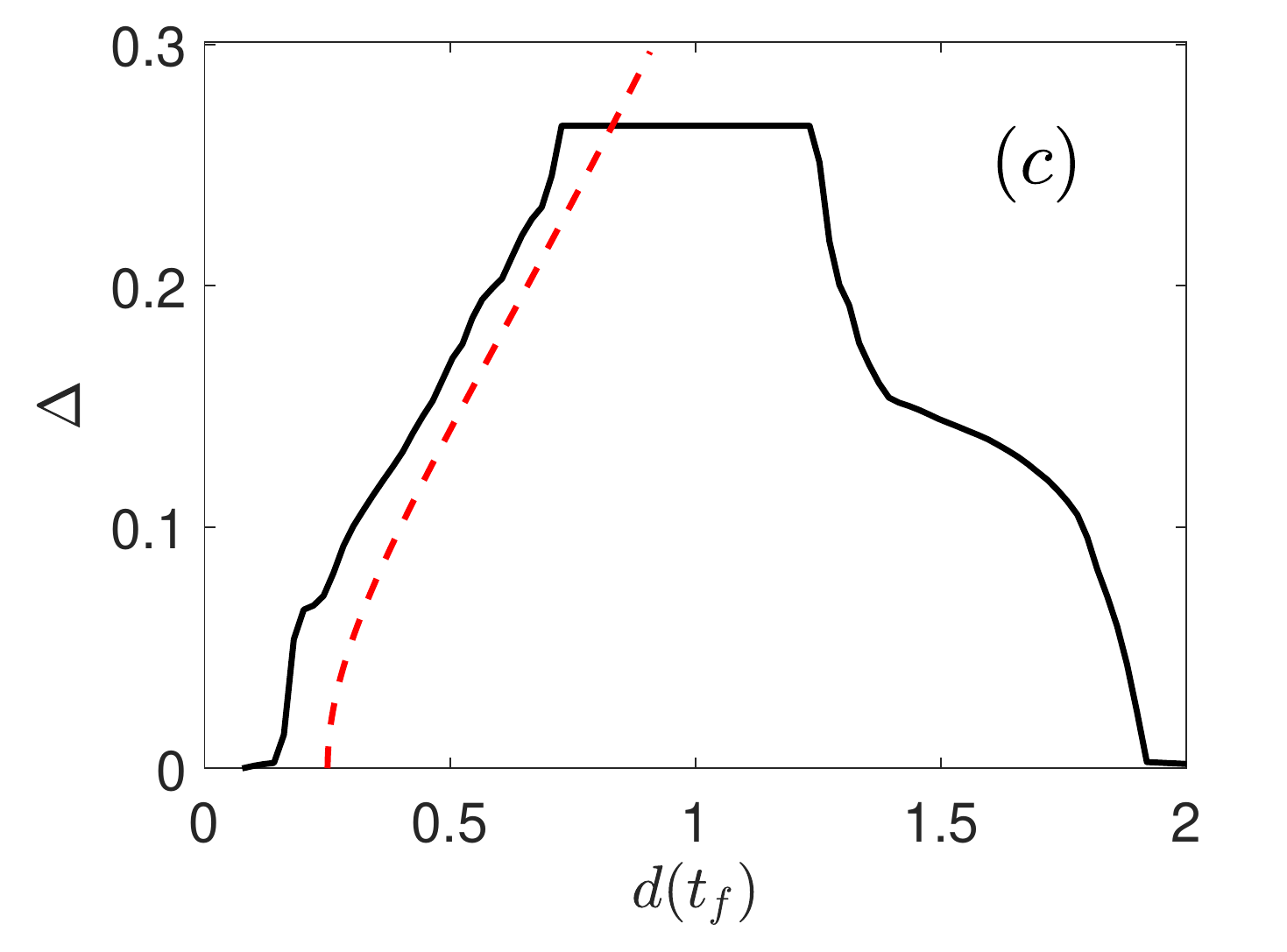}
   \caption{Comparison between the exact numerical computation (dashed red lines) and the estimations of the NN (black solid lines) for the direct (panel (\textbf{a})) and the inverse problems (panels (\textbf{b},\textbf{c})). In panel (\textbf{b}), the horizontal solid line indicates the maximum value of the offset term in the data set.\label{fig3}}
  \end{figure}

In order to interpret the results of Tables~\ref{tab:dataset12} and~\ref{tab:dataset34}, we compare in Figure~\ref{fig3} for different controls $u(t)$ the prediction of the NN and the results of an exact numerical computation both in the direct and inverse problems. Two different direct estimations are investigated in Figure~\ref{fig3}a. Similar results are achieved for other control protocols. The reasonable match between the two curves confirms that the NN can predict with a good precision the distance to the target state for any offset $\Delta\in [0,1]$ and any bang-bang control law with five switches in a fixed control time. This is a remarkable achievement for this global system characterization as no prior knowledge about the output or the control is required. Even if this result must be confirmed for more complex systems or different dynamics, no limitation was observed and it is a promising result with respect to the use of SL in quantum control. As could be expected from Table~\ref{tab:dataset34}, the conclusion is not so satisfying for the inverse problem. Differences are clearly visible between the exact computation and the prediction of the NN for some ranges of $\Delta$- values. A severe mismatch appears for a function $d(\Delta)$ which is not injective. Indeed, there are in this case values of $d$ associated with more than one offset $\Delta$. These multiple solutions cannot be found by the NN which gives only one value of $\Delta$ for each $d$. As can be clearly seen in Figure~\ref{fig3}b, the NN fails to predict the correct offset when the concavity of the curve changes, i.e., when the function is no longer monotonic. In Figure~\ref{fig3}c, the errors occur because of the non-monotonic behavior this time of the function $\Delta(d)$. It seems that a good match can be achieved if the function $d(\Delta)$ is always decreasing or increasing as shown by one of the examples in Figure~\ref{fig3}b. Due to the non-linear nature of the dynamics, this injectivity property cannot be determined a priori, making the application of SL more difficult in particular for global estimation issue. Note that this monotonic behavior of the mapping could be expected in a local analysis around a specific point $(\Delta,d)$. Finally, we also observe in Figure~\ref{fig3}b that the NN can predict offset values outside the selected variation range, such values are not physical and do not correspond to any quantum dynamic. This result is not very surprising because the NN was not guided or constrained at the time of its design.

Another interesting point concerns the executation time. Once a NN is trained, the inference time, which is required to estimate the predicted value, is smaller on a powerful GPU. In Table~\ref{tab:experiments}, the inference time is compared for all the NN architectures on a CPU, on a GPU of a laptop, and on a powerful GPU. The CPU is an Intel(R) Core(TM) i9-10885H CPU @ 2.40 GHz, one GPU is Quadro T2000 Mobile (this is a mobile GPU), and the other one is a Volta A100 GPU (currently the most powerful GPU). It is also instructive to make a comparison with a direct computation in Python of the dynamics. The computation of a data set with 10 million  systems takes 6587.30 s. So for only one element, it corresponds to a time of $6.59\times10^{-4}$ s. It can be concluded that there is an acceleration of magnitude 2 using a powerful GPU (100 times faster).

\begin{table}[tb]
\caption{Execution 
 times in second for the inference on CPU and GPU.}
    \centering
    \begin{tabular}{cccc}
    \hline
    \textbf{Algorithm}& \textbf{CPU} & \textbf{GPU T2000} & \textbf{GPU A100}\\
    \hline
    \hline
    MLP     & $1.36\times10^{-3}$ & $4.64\times10^{-5}$ & $6.81\times10^{-6}$ \\
    \hline
    CNN     & $4.31\times10^{-3}$ & $1.32\times10^{-4}$ & $9.31\times10^{-6}$\\
    \hline
    LSTM     & $3.05\times10^{-3}$ &$1.21\times10^{-4}$ & $4.36\times10^{-6}$\\

    \hline
    \end{tabular}

    \label{tab:experiments}
\end{table}

\section{Conclusions}\label{sec5}
We have applied recent formulations of SL techniques to the characterization of
two-level quantum systems by an external electromagnetic field. The model system under study can be viewed as an illustrative example to highlight the efficiency and the limits of this approach. The time evolution of the dynamical system is used to define different mappings in a direct or inverse way between the control law, the final distance to the target state, and the offset. The parameters of a NN are then optimized in a training process to approximate this mapping as accurately as possible. As described in this study, flexibility exists to fix the exact structure of the NN. Basically network complexity should reflect data complexity, but in practice only an empirical answer can be given with a trial and error method. A key and intrinsic difficulty  of the estimation procedures is the fact that they are global with no prior knowledge about the solution. Very good results are obtained for the direct problem, while the inverse one requires specific properties for the mapping, such as its monotonic~behavior.

This preliminary study opens the way to other analyzes in the same direction. It will be interesting to verify the conclusions of this paper in other quantum systems. An option to improve the results of the system characterization is to guide the algorithm by providing it with information about the dynamics. This general idea has already been developed in machine learning as part of physics-informed neural networks, which could be interesting to test in quantum control. In this work, a closed quantum system has been considered. A next step could be to extend this analysis to open quantum systems for the characterization, e.g., of the parameters describing the interaction between the system and the environment.

\appendix
\section{Code for the Different NN Architectures}\label{appa}
We give in this paragraph the different codes with the Keras framework to generate the NN architectures used  in this study (1 march 2023).\\

\textbf{Code for the MLP}\\
model.add(Dense(101, input$\_$dim=101, kernel$\_$initializer='he$\_$uniform',activation='relu'))\\
model.add(Dense(202, kernel$\_$initializer='he$\_$uniform', activation='relu'))\\
model.add(Dense(404, kernel$\_$initializer='he$\_$uniform', activation='relu'))\\
model.add(Dense(808, kernel$\_$initializer='he$\_$uniform', activation='relu'))\\
model.add(Dense(1212, kernel$\_$initializer='he$\_$uniform', activation='relu'))\\
model.add(Dense(1616, kernel$\_$initializer='he$\_$uniform', activation='relu'))\\
model.add(Dense(1212, kernel$\_$initializer='he$\_$uniform', activation='relu'))\\
model.add(Dense(808, kernel$\_$initializer='he$\_$uniform', activation='relu'))\\
model.add(Dense(404, kernel$\_$initializer='he$\_$uniform', activation='relu'))\\
model.add(Dense(202, kernel$\_$initializer='he$\_$uniform', activation='relu'))\\
model.add(Dense(101, kernel$\_$initializer='he$\_$uniform', activation='relu'))\\
model.add(Dense(50, kernel$\_$initializer='he$\_$uniform', activation='relu'))\\
model.add(Dense(25, kernel$\_$initializer='he$\_$uniform', activation='relu'))\\
model.add(Dense(1, kernel$\_$initializer='he$\_$uniform', activation='linear'))\\

\textbf{Code for the CNN}\\
model.add(Convolution1D(filters=8, kernel$\_$size=3, padding='same', activation='relu', input$\_$shape=(101,1)))\\
model.add(Convolution1D(filters=16, kernel$\_$size=3, padding='same', activation='relu'))\\
model.add(Convolution1D(filters=32, kernel$\_$size=3, padding='same', activation='relu'))\\
model.add(Convolution1D(filters=64, kernel$\_$size=3, padding='same', activation='relu'))\\
model.add(Convolution1D(filters=128, kernel$\_$size=3, padding='same', activation='relu'))\\
model.add(Convolution1D(filters=256, kernel$\_$size=3, padding='same', activation='relu'))\\
model.add(Convolution1D(filters=512, kernel$\_$size=3, padding='same', activation='relu'))\\
model.add(Flatten())\\
model.add(Dense(512, kernel$\_$initializer='he$\_$uniform', activation='relu'))\\
model.add(Dense(256, kernel$\_$initializer='he$\_$uniform', activation='relu'))\\
model.add(Dense(128, kernel$\_$initializer='he$\_$uniform', activation='relu'))\\
model.add(Dense(64, kernel$\_$initializer='he$\_$uniform', activation='relu'))\\
model.add(Dense(32, kernel$\_$initializer='he$\_$uniform', activation='relu'))\\
model.add(Dense(16, kernel$\_$initializer='he$\_$uniform', activation='relu'))\\
model.add(Dense(1, kernel$\_$initializer='he$\_$uniform', activation='linear'))\\

\textbf{Code for the LSTM}\\
model.add(LSTM(4*32, return$\_$sequences=True, input$\_$shape=(101,1)))\\
model.add(LSTM(4*32, return$\_$sequences=True))\\
model.add(LSTM(2*32, return$\_$sequences=True))\\
model.add(LSTM(2*16))\\
model.add(Dense(2*10, kernel$\_$initializer='he$\_$uniform', activation='relu'))\\
model.add(Dense(1, kernel$\_$initializer='he$\_$uniform', activation='linear'))\\

\end{document}